\documentstyle[12pt]{article} 
\begin{document} 
\setlength{\textwidth}{15 cm} 
\setlength{\textheight}{23cm}   
\baselineskip=18pt 
\bibliographystyle{unstrap} 
\renewcommand {\thefootnote} {\fnsymbol{footnote}} 
\newcommand{\figrtdwt}{1(a)~} 
\newcommand{\figrtlnqqt}{1(b)~} 
\newcommand{\figrtslw}{2~} 
\newcommand{\figrtquad}{3~} 
\newcommand{\figisoch}{4~} 
\newcommand{\figrtjsq}{5~}

\newcommand{\tabrttv}{1~} 
\newcommand{\tabisoch}{2~} 
\parskip=0.5cm 
\baselineskip=24pt 

\baselineskip=16pt 

\begin{center}

{\Large \bf Re-entrant Behavior of Relaxation Time with Viscosity at Varying
Composition in Binary Mixtures }
 
{\large \bf  Arnab Mukherjee, Goundla Srinivas and 
Biman Bagchi \footnote[1] {For correspondence: bbagchi@sscu.iisc.ernet.in}}\\

{\large Solid State and Structural Chemistry Unit \\
	Indian Institute of Science\\
	Bangalore -12, India\\}

\end{center}
\begin{center}
{\large \bf Abstract}
\end{center}
{\it In order to understand the long known anomalies in the
composition dependence of diffusion and 
viscosity of binary mixtures, we introduce here two new models
and carry out extensive molecular dynamics simulations.
In these models, the two molecular 
species ($A$ and $B$) have the same diameter and mass.
In model I 
the inter-species interaction is more attractive 
than that between the pure components,
while the reverse is true for model II.
Simulations and also mode coupling theory calculations
reveal that the models 
can capture a wide variety of behavior observed in experiments, most 
interesting among them are the non-monotonic variation of 
diffusion and viscosity with the composition and the {\em re-entrant} viscosity 
dependence of the relaxation time.}\\

{\em PACS} No:66.10.-x,66.20.+d,05.40.+j


 Transport properties of binary mixtures often show strong 
 and baffling dependencies on 
 the composition which have not been understood or even adequately
 addressed to in a molecular theory.  
 The well-known Raoult's law of classical physical chemistry$^1$ 
 predicts the following simple linear dependence on the composition
 for a given property $P$, 
\begin {equation}
P = x_1 P_1 + x_2 P_2
\end {equation}
\noindent where $x_i$s are the mole fractions and $P_i$s are the 
 values of the property $P$ of the  pure (single component) liquids.
 More often than not, significant deviation from Eq. 1 is observed.
 Of many anomalies exhibited by binary mixtures, the existence of
 an extremum (sometimes even a double extrema!) in the composition 
 dependence of excess viscosity$^2$  and the re-entrant
 type behavior of the relaxation time when plotted against viscosity$^3$
 are certainly the most remarkable. The latter shows, in a dramatic fashion, 
 that {\em viscosity is not a unique determinant of relaxation in binary
 mixtures}$^3$. 
 Several interesting theoretical and  computer simulation studies 
 on Lennard-Jones binary mixtures have been carried out recently
 $^{4-6}$, but these studies have 
 concentrated mainly on the glass transition in binary mixtures and 
 considered {\em only one particular composition}
 and a unique interaction strength.
 Earlier Heyes carried out the extensive equilibrium MD simulations
 of Lennard-Jones binary mixtures by using both the microcanonical 
(N V E) and canonical (N V T) ensemble methods to study the partial 
 properties of transport coefficients in the inert gas medium $^{7}$.
 The non-ideality in the case of inert gas 
 mixtures is small, since their mutual interaction strength 
 ($\epsilon_{AB}$) follows the Berthelot mixing rule.  

 In order to understand the markedly nonlinear composition dependence, 
 here we introduce and study 
 two models (referred to as model I and model II) of binary mixtures in which
 the solute-solvent interaction strength 
 is varied by keeping {\em all the other parameters} unchanged. 
  In our models, all the
 three interactions (solute-solute, solvent-solvent and solute-solvent )
 are described by the $(6-12)$ Lennard-Jones potential,
\begin {equation}
U_{ij} = 4 \epsilon _{ij} \left [{\left (\sigma \over r_{ij}\right )}^{12} - 
{\left (\sigma \over r_{ij} \right )}^6 \right ]
\end {equation}
\noindent where $i$ and $j$ denote any two different particles. 
We set the diameter ($\sigma$) and mass ($m$) of both the solute and 
the solvent molecule to unity, for simplicity. The solvent-solute interaction 
strength lies in the potential well depth 
$\epsilon_{AB}$, where $A$ and $B$ represent the solvent and solute
particles, respectively. Throughout this study we keep  
the interaction strength $\epsilon_{AA}=1.0$, (solvent-solvent), 
$\epsilon_{BB}=0.5$ (solute-solute). 
In the two models  we use two different solvent-solute 
interaction strength values, namely 
$\epsilon_{AB}=2.0$ in model I and  $\epsilon_{AB}=0.3$ in model II. 
So, while model I is a "structure former"
(between $A$ and $B$), model II is a "structure breaker". Note that both
$A$ and $B$ have the same diameter. We believe that these simple models can
serve as starting points to understand the many baffling properties 
of binary mixtures.

  Extensive MD (microcanonical 
ensemble, with the usual periodic boundary condition)
simulations have been carried out with a total of $500$ particles
for two types of models by varying the solute
 mole fraction (that is, of $B$) from 0 to 1.
The reduced temperature $T^*$ ($=k_B T/\epsilon$) is set equal to
unity in model I and 1.24 in model II and the reduced density 
$(\rho^*=\rho \sigma^3)$ is 0.85 in both the models. After many
trial runs to verify the existing results on viscosity $^{7}$ 
of one component liquids, we have selected 
a time step $\Delta t^*=0.002 \tau$ for model I and $\Delta t^*=0.001 \tau$ 
for model II. where, 
$\tau = \sigma \sqrt{m/\epsilon}$. We have dealt with six different
solute compositions, namely 0, 0.2, 0.4, 0.6, 0.8 and 1.0. For each 
solute composition we have equilibrated the system up to $1.5 \times 10^5$ 
steps. Simulations carried out for  another $2.0\times 10^5$ steps
after the equilibration during which all the relevant quantities 
have been 
calculated. For each composition, we have run three independent simulations
and have taken averages over them. 
We have checked all the three partial radial distribution 
functions to make sure no phase separation occurs during simulations
(for model II). 
 Viscosity values are obtained
by integrating the stress time correlation function which defines the time
dependent viscosity by the following relation
\begin{equation}
\eta(t) = (Vk_BT)^{-1}\langle\sigma^{xz}(0)\sigma^{xz}(t)\rangle
\end{equation}\\
where the off-diagonal element of the stress tensor $\sigma^{xz}$ for binary
mixture is defined as,
\begin{eqnarray}
\sigma^{xz} = \sum_{j=1}^{N_{1}}[(p_{j}^{x}p_{j}^{z}/m)+ F_{j}^{z}x_{j}]
  +  \sum_{j=N_{1}+1}^{N}[(p_{j}^{x}p_{j}^{z}/m)+ F_{j}^{z}x_{j}]
\end{eqnarray}  
Here, $F_{j}^{z}$ is the z-component of the force acting on the $j$-th particle
and the corresponding X-coordinate is $x_j$, $p_{j}^{z}$ is the z-component 
of the
momentum ${\bf p_j}$ of $j$-th particle, $m$ being the mass of the particle. 
Amongst total 
$N$ particles, $N_{1}$ solvent particles are labelled from  $1$ to $N_{1}$ 
and solute particles from $(N_{1}+1)$ to $N$.                    

 Diffusion coefficients are calculated both from the mean square displacement
 and velocity autocorrelation function via the Green-Kubo relation. 
 The results of the simulation are given in figures $1-4$. We shall discuss
 the results after we describe the mode coupling theory employed.

 We have carried out mode coupling theory (MCT) calculations of diffusion
 and viscosity to understand
 the simulation results, especially the origin of non-monotonicity.
 These calculations  have been carried out by using well-established expressions
 $^{8-12}$. Note that for binary mixtures at normal density and 
 temperature, the short
 time dynamics of the relevant time correlation functions are important 
 and in fact, can contribute more than 50{\%} of the total value, just as
for one component liquids. Thus, any
 solution of the MCT equations require accurate input
 of the short time dynamics. 
 For a given transport property $P$, MCT formalism$^{8-12}$ assumes 
 the following
 separation into the short time, binary collision controlled, contribution
 $P^{(bin)}(t)$ and the contribution from the collective term, which in dense 
 liquid
 is dominated by the density term, $P^{(\rho\rho)}(t)$. So the total dynamical 
 quantity $P(t)$ can be written as $^{8}$,
\begin{equation}
P(t)= P^{(bin)} (t) + P^{(\rho\rho)}(t)
\end{equation}
The binary part of  both viscosity and the friction are assumed to be
  Gaussian, $^{8-10}$  
\begin{equation}
 P^{(bin)}(t) = P(t=0)exp(-t^2/{\tau_{P}^2}).
\end{equation}
\noindent For viscosity, $P(t=0)$ is the high frequency shear 
modulus $G_{\infty}$ of a binary  mixture given by, $^{15}$
\begin{equation}
G_\infty = (\rho_1 + \rho_2)k_BT + {\frac{2\pi}{15}}\sum_{i,j=1}^2{\rho_i}
{\rho_j}\int\limits_{0}^{\infty}dr g_{ij}(r)\frac{d}{dr}
\left [ r^4\frac{dv_{ij}(r)}{dr}\right ]
\end{equation}
\noindent here, $i,j=1$ indicate solvent particles and 
$i,j = 2$ denote solute 
particles. $\rho_{1}$ and $\rho_{2}$ are the coresponding number densities for
the solvent and solute particles. $g_{ij}(r)$ is the partial radial
distribution function of the particles labelled $i$ and $j$. In the present
calculation, $g_{ij}(r)$ is obtained from Ornstein-Zernike equations with SMSA 
closure$^{14}$,
which provides a reasonable, although not perfect, agreement with the simulation
results. 

 The characteristic relaxation time $\tau_{\eta}$ for viscosity
can be determined by the second derivative of $\eta(t)$.
The resulting expression of ${\tau_{\eta}}$ for a binary mixture 
is rather complex and is given elsewhere $^{15}$.
The mode coupling contribution to viscosity is assumed to be given by
the binary product of the density terms$^{8,9}$.
In the present case, one derives contribution from four such density
terms and total MCT term $\eta^{(\rho\rho)}$ is given by the simple addition of 
four $\eta^{(\rho_{i}\rho_{j})}$ terms.
The frequency dependent diffusion coefficients 
 $D_{i}(z)$ are related to the respective frequency dependent frictions according to
 Einstein relation,
\begin{equation}
D_{i}(z) = C_{vi}(z) ={{k_{B}T}\over{m_{i}(z+\zeta_{i}(z))}}
\end{equation}
\noindent For friction, $\zeta_{i}(t=0)$ is the Einstein frequency of 
the $i$-th component in the mixture and is 
determined by the static correlation
functions. The initial short time part of time dependent 
friction is assumed to be Gaussian with characteristic time 
$\tau_{\zeta}$ which is calculated using the short time expansion of
the force-force time correlation function $^{8-10}$.
The expression for the mode coupling term has already been given by 
Bosse {\it et al} $^{11}$ and need not be presented here.
The dynamical input parameters for MCT calculation are the 
wavenumber ({\bf q}) dependent and time ($t$) dependent partial intermediate
 scattering functions, $F_{ij}(q,t)$ and the self-dynamic structure
 factors, $F_{si}(q,t)$.
 The expressions of dynamic structure factors $F_{ij}(q,z)$ are obtained by 
 solving four coupled equations obtained from time dependent 
 density functional theory $^{13}$. This method requires the value of
 the frequency dependent self-diffusion coefficient of both the species. 
 We have used the bare
 value of the diffusion coefficient here, calculated from  Eq. 8 by using 
 the binary part
 of the frequency dependent friction.
  We approximate $F_{si}(q,t)$ by 
$F_{si}(q,t)=exp({{-q^{2}\langle\Delta r_i^{2}(t)\rangle}\over {6}})$
where
\begin{equation}
\langle \Delta r_i^{2}(t)\rangle = 2\int_{0}^{t}C_{vi}(\tau)(t-\tau)d \tau
\end{equation}
$C_{vi}(t)$  in Eq. 9 is obtained from the numerical inverse Laplace transform 
of $C_{v}(z)$. 
  
  
  Figure $1$ shows the remarkable
  re-entrant behavior of the structural relaxation 
  times ($\tau_{i}$) when the viscosity is changed by varying the composition.
  The increase in composition is indicated by arrows. Here the relaxation 
  time $\tau_{i}$ is calculated by 
  using $\tau_{i} = \sigma^{2}/D_{i}$. Thus, the relaxation time is
  inversely proportional to diffusion constant. We believe that a
  similar behavior will be observed for rotational relaxation
  as well. The simulation points here are averages over three 
  independent long runs;
  error bars are typically $\pm 0.2$ for viscosity and $\pm 1.5$ for
  the relaxation time (that is, $\pm 0.002$ for diffusion coefficients). 
   Note that in this figure we have shown only
  the simulation results, for clarity -- theory shows a similar behavior. 
      
   Figure $2$ shows the composition dependence of diffusion 
  coefficients obtained from both theory and 
  simulation, for model I. Figure $3$ shows the same for model II.
  Note the non-monotoic composition dependence. Diffusion of $A$ and $B$ show
  differing behavior, in all the cases.
  Figure $4$ depicts the nonideality
  of viscosity with respect to composition, for both the models.
  Though the agreement between theory and simulation is certainly not perfect, 
  the trends  are similar in both the calculations. Note that the theoretical 
  calculation does not use {\em any} simulation data as input or 
  {\em any adjustable 
  parameter} either; thus the theory and the
  simulation provide independent test of each other which is important for 
  binary mixtures.
  
    We conclude this Letter with the following comments.

 (1) It is shown that the simple models can describe the
 decoupling of diffusion from viscosity in binary mixtures, when the 
 viscosity is changed by varying the composition. This
 decoupling is most dramatically manifested in the re-entrant type behavior
 depicted in Fig. 1. This
 shows that {\em viscosity is not a unique deteminant of relaxation in binary
 mixtures.}
  
 (2) Models I and II seem to reproduce the behavior observed in large
  number of systems. We believe that this is the first time a microscopic
  model captures the strong non-ideality of diffusion and viscosity. The 
  results agree with the age
  old wisdom that structure making interactions between the two constituents
  (here $A$ and $B$) lead to a slower relaxation.
  The opposite has also been observed
  for model II which has the structure breaking interactions.
  
 (3) What is also remarkable is that the non-ideality manifests itself
  in nearly opposite ways for the two models. This has been observed
  for all the transport properties 
  and hence reflected  in  the re-entrance behavior also. 
  
 (4) The theoretical calculations reveal that the main reason for the 
 anomalous composition dependence of viscosity
 lies in the variation of the mean 
  square stress fluctuation (MSSF) with the composition of the mixture.
  Similarly, for friction,  it is the Einstein frequency which shows
  non-monotonic behavior. 
  It is thus fair to say that the anomalies have both a structural and
  a dynamic  origin.

 This work was supported in parts by the Department of Science and
 Technology, India and CSIR, New Delhi, India. 
 G. Srinivas thanks CSIR for a research fellowship. 


\newpage

{\large \bf Figure Captions}

{\bf Figure 1} The simulated values showing the re-entrant behavior 
of the relaxation times $\tau_{i}$, 
are plotted against simulated viscosity for model I. Filled circles 
represent $\tau_{A}$ while the open circles $\tau_{B}$.
The direction of the arrow shows the
increasing solute ($B$) composition in both the cases. $T^{*}=1.0$,
$\rho^{*}=0.85$.

{\bf Figure 2} The diffusion coefficients obtained from MD simulation and
mode coupling theory are plotted for model I. 
Filled and 
open cirles represent the solvent and solute diffusion coefficients
obtinaed from simulations, respectively. Full and dashed lines
show the MCT results. $T^{*}=1.0$, $\rho^{*}=0.85$.

{\bf Figure 3} The diffusion coefficients obtained from MD simulation and
mode coupling theory are plotted for model II. 
Filled and 
open cirles represent the solvent and solute diffusion coefficients
obtained from simulations, respectively. Full and dashed lines
show the MCT results. 
$T^{*}=1.24$, $\rho^{*}=0.85$.

{\bf Figure 4} The composition dependence of viscosity obtained from
MD simulations (symbols) and  mode coupling theory (lines)
for both the models. Filled (open) circles give simulation results
for model I (model II). The lines give the theories.

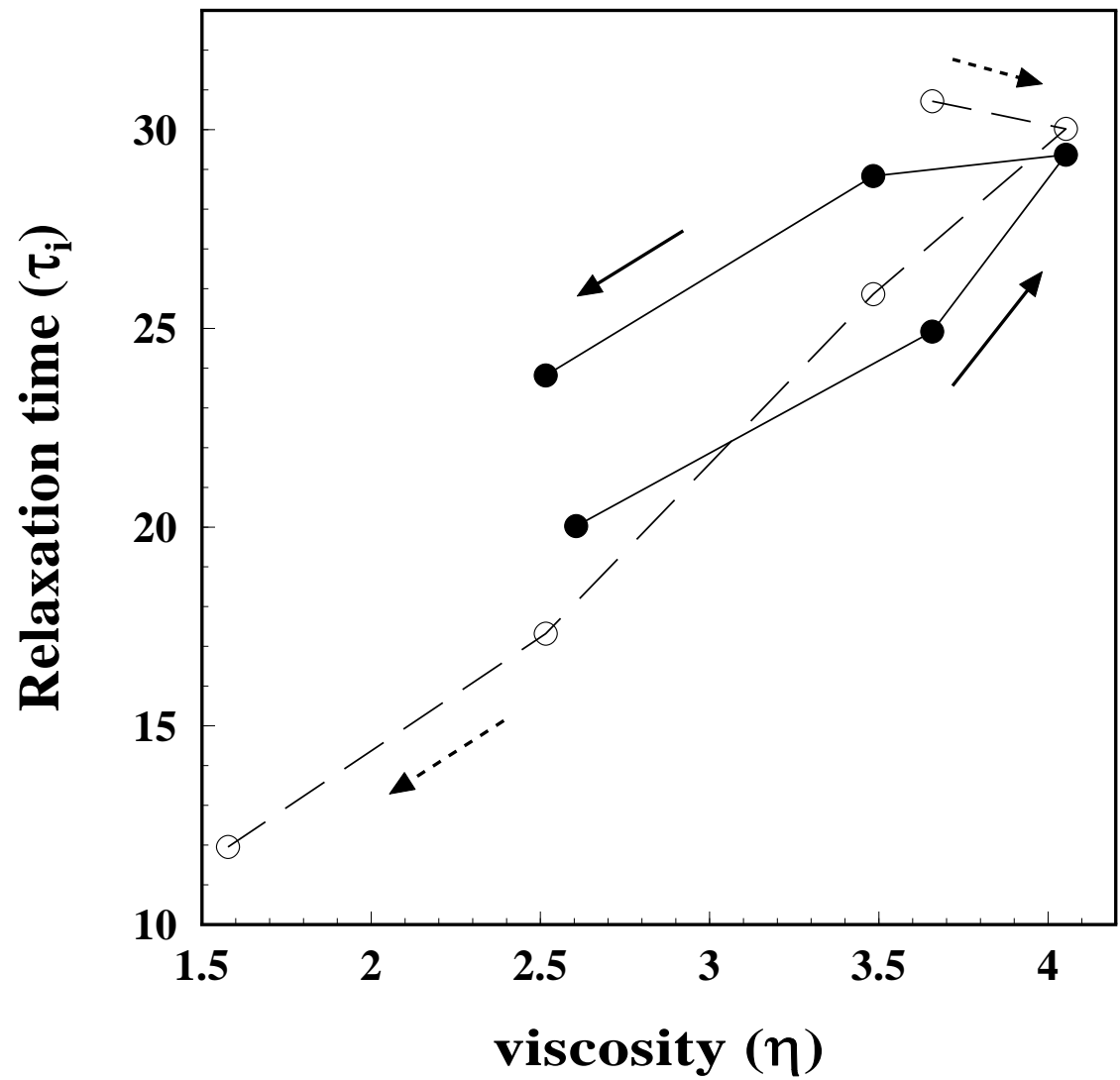
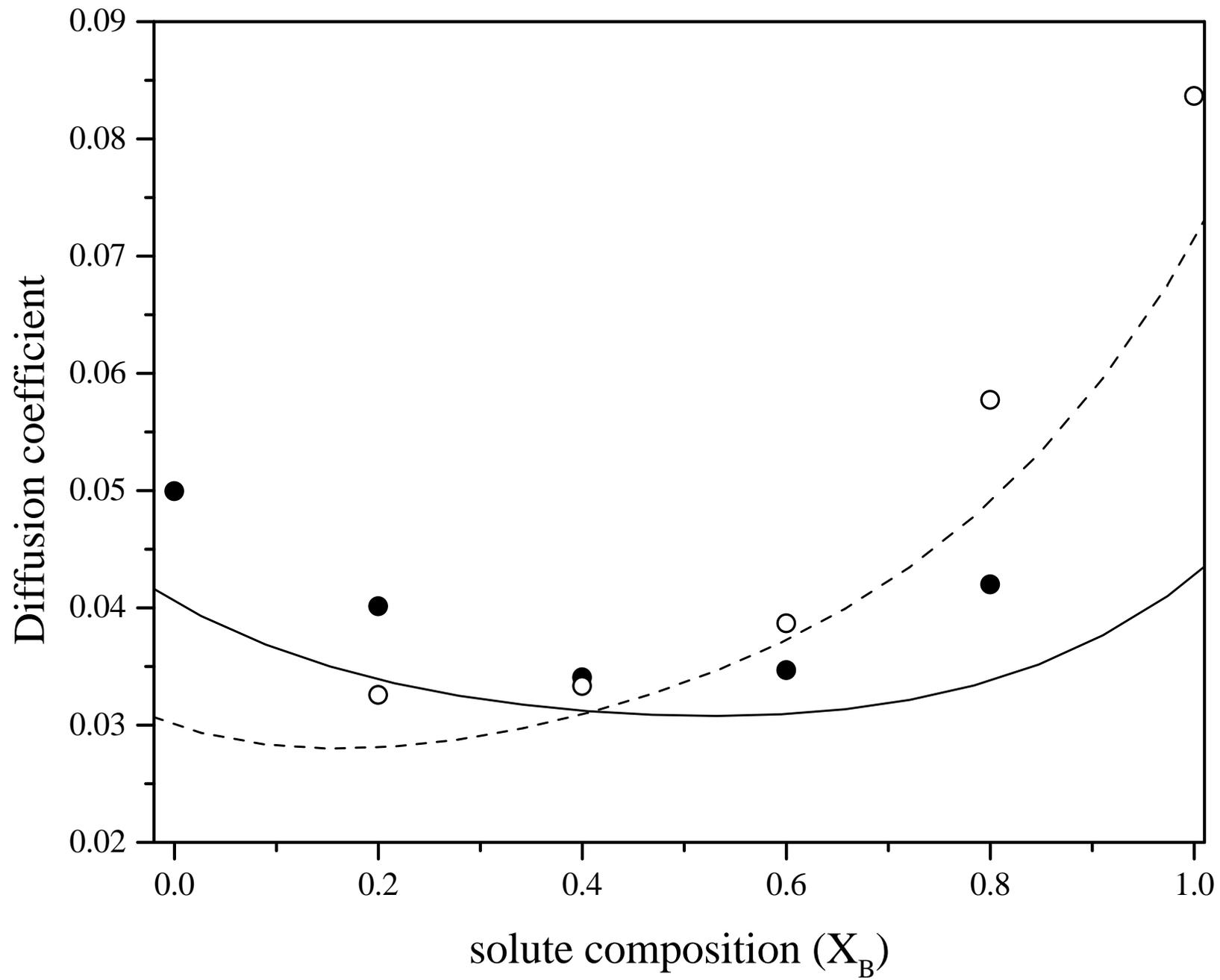
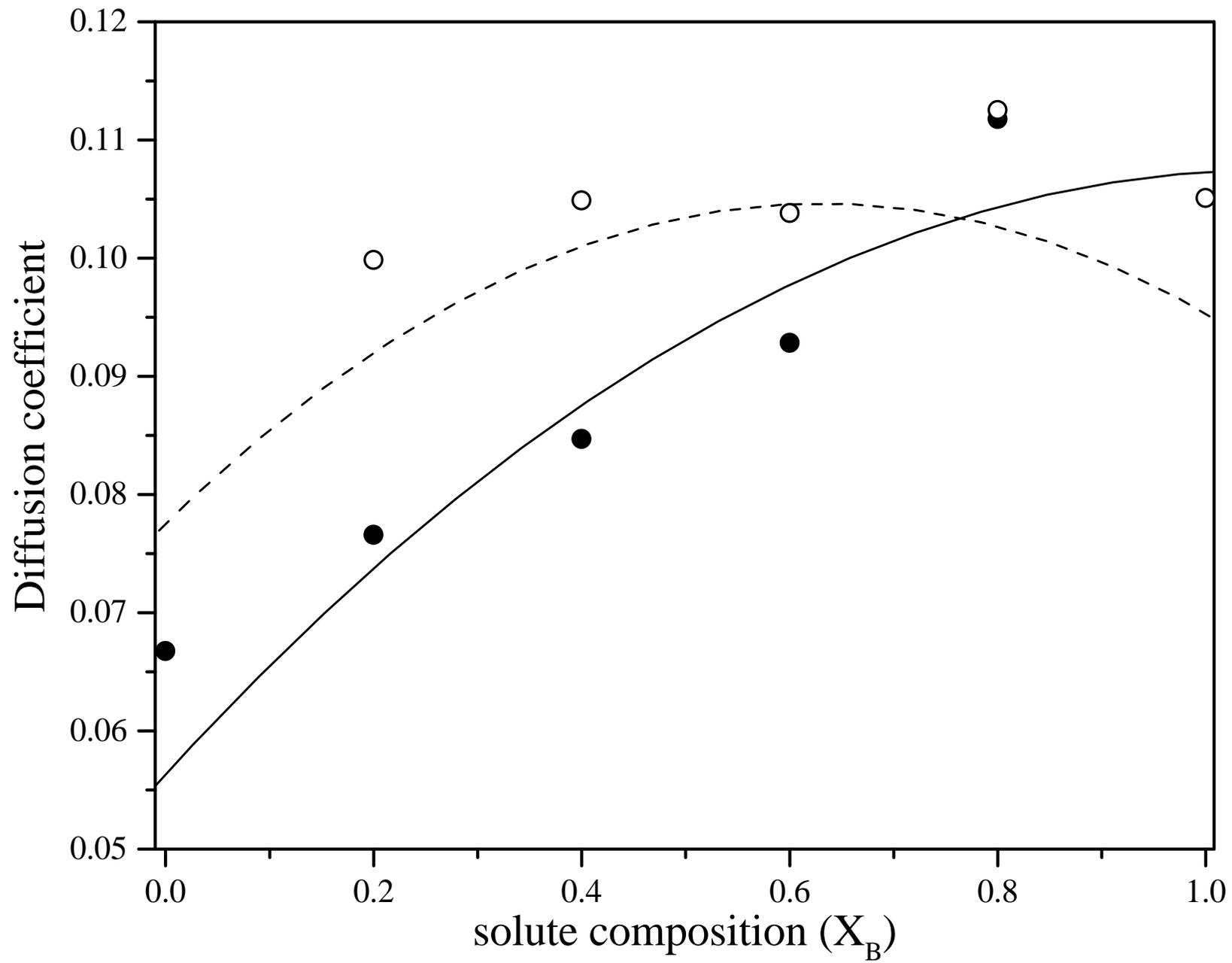
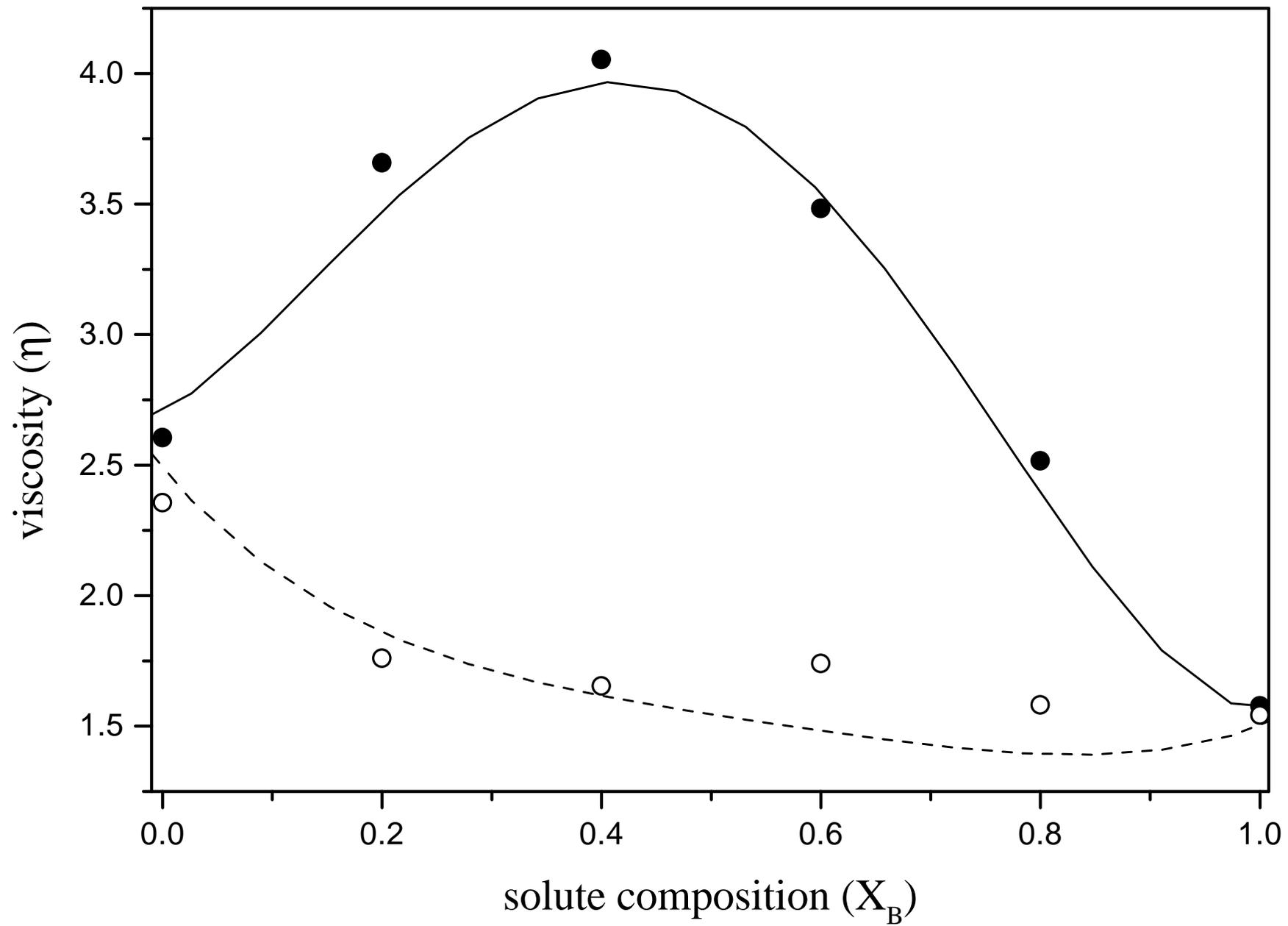
\end{document}